\documentstyle[12pt,version2,aps]{revtex}
\begin{document}
\renewcommand{\theequation}{\arabic{section}.\arabic{equation}}
\newcommand{\eqreset}{\setcounter{equation}{0}}
\setlength{\textheight}{20cm}
\vspace*{.9 in}
\begin{center}
{\large\bf ENHANCEMENT of
 PERSISTENT CURRENT ON MULTICHANNEL RING$^*$}

\vspace{.5 in}
{\sc F.V. Kusmartsev}

 \vspace{.3 in}

{\it L.D. Landau Institute for Theoretical Physics \par
Moscow,117940, GSP-1, Kosygina 2, V-334, Russia \par }

and

 \vspace{.3 in}
     {\it Institute for Solid State Physics,University of Tokyo}\\
        {\it Roppongi,Minato-ku,Tokyo 106 } \\

(received ~~~~~~~~~~~~~~~~~~  )
\end{center}

 \vspace{.9 in}

*)submitted to Physical Review Letters
\vfill
\eject
\begin{abstract}
We study the effect of strong electron-electron interactions
on the persistent current in a multichannel ring with the aid of the Bethe
ansatz equation discussed by Sutherland \cite{Suth} and Tsvelik\cite{Tsve}. The
interaction acts  differently
 on a single and multichannel ring. If, in the first case, the
current is suppressed, then in the second case the current may be enhanced.
With the interaction the current is proportional to square of the number of
channels. The maximum effect corresponds to medium values of the interaction,
where $U\sim V_F$.
The new phenomenon of the fractional $p/q$ Aharonov-Bohm effect
appears for interacting electrons on the multichannel ring.
\end{abstract}

In 1959 Aharonov and Bohm \cite{Ahar}
 proposed an experiment, where they described the
periodic dependence of an
electron beam difraction pattern   on the
magnetic flux of the tiny
solenoid penetrating the beam. One year later such
single flux quantum periodic dependence was observed.
 There is also a single flux quantum periodic dependence
 for a
persistent current  in a small metallic ring, penetrated
by a  magnetic flux, similar to that in the Aharonov-Bohm experiment, which is
also  known as  Aharonov-Bohm (AB) effect \cite{Land},\cite {Kuli}
-\cite {Liu}.

Now, let us discuss the role of an electron-electron interaction in
the AB effect.
First of all, we consider the single channel ring with
spinless fermions in a tight-binding model, where there is
 only a repulsion between electrons located on  next - neighbor sites.
The numerical simulation indicates that there is no much difference between
short and long-range interaction\cite{Mull}.
It is intuitively clear that
the interaction induces a repulsion between levels.
The latter causes  the Fermi velocity $V_F$  and hence
the persistent current to be reduced.
However, let us note that the continuum and lattice model give different
results\cite{Mull}. For the continuum single channel model the two-particle
interaction does not
change the energy-flux dependence \cite{Mull}, which is in a contrast
with the lattice models\cite{Kus}.

In  weak coupling this effect
may be shown  with the aid of
a bosonisation in the tight-binding model, where there is
the next neighbor repulsion $W$.
As a result, in the expression for the persistent
current for free fermions:
\begin{equation}
J_P \sim -\frac{V_F}{L} (f-p)
\end{equation}
we must only change the Fermi velocity
\begin{equation}
V_F \rightarrow
 U_F={V_F} \sqrt{1-\frac{W}{V_F}}
\end{equation}
where the parameter $p$ is an integer or half-integer number( we use the units
$e=\hbar=c=1$). The current
 depends on the
parity of the total number of fermions $N$ \cite {Kus},\cite{Legg},\cite
{Loss}.
However, in the continuum model the Fermi
velocity is unchanged\cite {Loss}.

Thus, at small $W/V_F$ the suppression of the current is almost
linear in $W$. However, with  increasing  $W$, the suppression
is saturated and the current amplitude takes a limiting value depending
on the number of fermions on the ring. Applying the
Bethe ansatz to the limit $W/V_F\rightarrow\infty$, we have the following
exact expression for the current \cite {Kus}:
\begin{equation}
J_p=   -\frac{4\pi}{L} \frac{\sin\frac{\pi N}{L-N}}{\sin\frac{\pi}{L-N}}
       \sin\frac{2\pi}{L} (f-p)
\end{equation}

The impurities on a single channel ring induce a repulsion between levels,
therefore one may expect the suppression of the persistent current.
However, the interaction may effectively renormalize the
localization length, which slightly increases, i.e.
there might be a little
enhancement (
see, for comparison, Ref.\cite{Abra}\cite{Mull}\cite{Weis}\cite {Bouz}).

Thus for a single channel ring 1) there occurs a  suppression of the persistent
current;
2) with the disorder the suppression is slightly reduced or enhanced.
3) there exists a parity effect: a difference in the current
behavior for an even and an odd number of particles
 \cite {Kus},\cite{Legg},\cite {Loss}.

However, the situation is drastically changed when there are two
or many channels or even if
we take into account the spin quantum numbers. To show this let us consider
 a  ring consisting of quantum dots. The system has effectively
two channels for up-spin and down-spin particles,
 and is described by  Hubbard model with $t$
 as a hopping element between dots, and
 $U$ as a charging energy $U={e^2}/{C}$, where
$C$ is the capacitance or  radius of a single dot.
Then,  the Hamiltonian reads
\begin{equation}
H = -t \sum_{i,\sigma} C_{i,\sigma}^+ C_{i+1,\sigma} + U \sum_i N_{i\uparrow}
N_{i\downarrow}
\end{equation}
where $U$  is a characteristic energy of a level splitting  on a single dot.

One may  strightforwardly estimate the persistent current in the weak
 coupling, i.e. when ${U}/{V_F}<< 1$ via bosonisation \cite {KusTreste}. Doing
so, we see that  in this  case the spin and charge degrees of freedom
are decouple, i.e.
there are spinons
and holons and
there is no suppression of the persistent current.
 The effective Fermi
velocity of charge degrees of freedom (holons), characterising the current,
 does not decrease, like that for spinless fermions, but,
instead, increases
\begin{equation}
V_F \rightarrow U_{CF}= V_F \sqrt{1+\frac{U}{V_F}}
\label{ren-fer-vel}
\end{equation}
In  contrast with the holon Fermi velocity,
the Fermi velocity of spinons decreases
$V_F \rightarrow U_{SF}=V_F \sqrt{1-\frac{U}{V_F}}$.

The persistent current is carried by holons, having a charge
but   no  spin. Note that to create a single
holon we have to have two electrons, as  follows from
 bosonisation.
For examle, a single holon is equal to an electron
 with up-spin $\uparrow$ plus
an electron with down-spin $\downarrow$, resulting in a charge without a spin (
a fluctuation of the charge density, only). Therefore, the Fermi momentum of
holons is $2 k_F$.
Therefore, the Aharonov-Bohm effect or persistent current is characterized by
half-flux quantum periodicity. Thus, the reason of such periodicity
is a decoupling of the spin and charge degrees of freedom.

To get the solution in the strong-coupling limit, we expand the Bethe equation
in the parameter
$\alpha={V_F}/{U} \cong  {tN}/ {UL}<<1$,
which is equivalent to a large ratio $U/t$ or a low density of electrons
$N/L<<1$. In the zeroth approximation the spectrum is
\cite{Kus}:

\begin{equation}
K_n = \frac{2\pi }{L}(n+ f+ \frac{\sum_{\alpha} J_{\alpha}}{N}),
\end{equation}
which is different from that
for spinless fermions, which is $K_n = \frac{2\pi }{L}(n+ f)$, by an additional
statistical flux
$\Phi{st}= {\sum_{\alpha} J_{\alpha}}/{N}$,
 where the $J_{\alpha}$ are the
quantum numbers of spinons. The statistical flux is changed
by fractions $l/N$, where $l$ are arbitrary integer numbers.
This means that the persistent current is a $1/N$ periodic
function\cite{Kus1} and   described by the formula:
\begin{equation}
J_P \sim -\frac{V_F}{L} (f-p)
\end{equation}
where  $ p=l/N$. In other words, the current displays the
fractional $\frac{1}{N}$ Aharonov-Bohm effect\cite{Kus1},\cite
{Scho},\cite{Yu}.
Note, that the     holes have multivalued wave function, which may be
related to fractional  statistics.

When
$N\rightarrow \infty$ the amplitude of
the fractional $ {1}/{N}$ oscillation of the current
vanishes, i.e. the fractional effect disappears.
Then, the main periodicity will  come from the first correction
in the parameter $\alpha$.
In spite of the fact that  the spin and charge degrees of freedom are decoupled
the spinons create a fictitious field $\phi_{fic}$,
in which the holons move,
\begin{equation}
\phi _{fic} = \frac{2\pi\sum J_{\alpha}}{N} = \sum_n k_n=K_s.
\end{equation}
where $K_s$ is the total momentum of spinons.

To estimate the dependence of the spinon energy on the
fictitious flux we will use the Bethe ansatz equations(see,\cite {KWKT} and
references therein).
First of all, let us consider the case when the spinon band is
half-filled, that is, when the number of up-spin  particles$N_{\uparrow}=M=
{N}/{2}$.

In the expansion in the parameter $\alpha$, the
 spinons are described effectively
 by a Heisenberg Hamiltonian\cite{KWKT}. The dependence of
the spinons' energy on the total momentum (fictitious flux) is
the energy of the spin-wave excitations, so we get
\begin{equation}
E_s = \frac{V_F \alpha}{L} \mid \sin \phi_{fic} \mid.
\end{equation}
The total energy of the Hubbard ring is the sum of
 the holon and the spinon energy:
\begin{equation}
E = \frac{V_E}{L} (f-\phi)^2 +\frac{V_F}{L} \alpha\mid \sin \phi\mid
\label{total}
\end{equation}
where fictitious flux takes the values
 $\phi = 0,\pm\frac{2\pi}{N}, ..$.

With a change of the flux $f$,
at zero fictitious flux $\phi$
 the holon energy
 (the first term in eq.(\ref {total})) increases rapidly.
However, if with the change of $f$ we also change $\phi$,
 exciting spin-waves, the contribution of the first
term may vanish and the total energy decreases.
Thus in the large $N$ limit,
 we must put $\phi=f$ and the ground state energy dependence
 is described by the second term in the eq.(\ref{total}).
This dependence is a $\frac{1}{2}$-flux quantum periodic function,
which is  an envelope of $\frac{1}{N}$-flux periodic parabolas.
Therefore, the persistent current is described by the
expression
\begin{equation}
J_p\sim -\frac{V_F}{L} \alpha  \cos f
\label{ev-cur}
\end{equation}
where $ 0<f<\pi $. There is also a
difference (a quarter flux quantum shift) in the behavior of
the persistent current for even  and for odd $ N $ (parity effect).

So, we arrive at the conclusion that
the structure of the spin-wave excitation spectrum,
i.e. the dependence on the spin wave momentum, determines the
 AB periodicity as well as the parity effect. Comparing the
formula (\ref{ren-fer-vel}) with the formulae(\ref{ev-cur})
one sees that the amplitude of the current is maximal
at intermediate values of $U$, when $U\sim V_F$. That is, on the
two channel ring there is an enhancement of the persistent current,
which is maximal
at intermediate values of the electron-electron interactions.

In the case of arbitrary $M$ we find that
the envelope function of ${1}/{N}$ flux quantum periodic
parabolic curves is an
${M}/{N}$ periodic harmonic (for details, see Ref.\cite{KusT}).
 Strictly speaking,  there are
${M}/{N}$ and single flux quantum periodic oscillations
of the persistent current. The $M/N$ oscillations survive even if
the $\alpha$ increases, see, Fig.1, where we show
the 1/3 periodic oscillations of the ground state energy flux dependence
( low energy parabolic-cusp curve).
 The reason of such periodicity is a cuspoidal shape
of the spin wave excitation spectrum, which is presented
by short horisontal  lines on this Figure. One sees that in this case the
persistent current
is determined by lowest parabolas and, therefore, its maximal value
is proportional to  $\sim 10 V_F/L$.

 However, when $\alpha$ descreases,
the energy of spinons decreases and at some flux value this energy
becomes the lowest one ( see, Fig.2A ). Then, in this region the $1/N$
oscillations are created
which envelope is determined via the spinon excitation energy.
Therefore, in this case instead of parabolic-cusp function (see, Fig.1)
the ground state energy-flux dependence becomes a smooth function
(see, Fig.2B).

The most stricking result, however,  is that the persistent current is  a
perfect $1/3$ periodic   function of the flux in units
of elementary flux quantum ( see, Fig. 3). There are two different regions,
where  the current
effectively increases and monotoneosly decreases. In the flux region
where the current effectively increases there occurs the perfect $1/N$ periodic
oscillations, which are practically not resolved on the scale of
 the Figure. The current
decreases linearly on the each $1/N$ flux region. The slope of this decrease
in the single $1/N$ oscillation is equal to the slope in the region where
the current monotoneously decreases (see, Fig.3). The maximal
current amplitude is equal to $\sim 10 V_F/L$. Note, that this large amplitude
of the current is determined via the small $1/N-$ oscillations, related
to the $N-$ particle bound state\cite{Kus1}. Therefore on the Hubbard
ring with the disorder the localization effects are weaker, since the
localization length of this $N-$ particle bound state is in $N-$ times larger.

Generally speaking, the
amplitude of the ${1}/{N}$ current oscillations
is proportional to $\sim{V_F}/{LN}$.
The amplitude of ${M}/{N}$  flux quantum
periodic current oscillations is proportional to
 $\sim {V_F\alpha}/{L}$.
Thus, the ${M}/{N}$ periodic current oscillations are
dominant if $\alpha>>{1}/{N}$ ( see, for details Ref.\cite{KusT}).

It is obvious that in the more complicated double chain ring, the
fractional ${M}/{N}$ periodic oscillations can occur,
provided that there is an effective separation of the "spin and charge"
degrees of  freedom. The "spinons", whose number equals $M$, are
 associated with
the relative fluctuations of the charge density and the "holon"
--- with the total fluctuation of the charge density in these two channels.
As a result, the total momentum of "spinons" is
a fictitious flux for the motion of the "holons". This fictitious flux
creates the fine structure of AB effect on the double chain ring.


 There is a generalization of the results obtained to
a multichannel ring.
So, let us assume that there are p-channels,
where there arises  a relative
splitting of degrees of freedom into $holons$ and $p-1$ types of "spinons".
The problem may be studied  with the aid of the Bethe ansatz
for the $SU(N)*SU(M)$ model. The Bethe ansatz equation for this model has been
discussed by Tsvelik \cite {Tsve}.
We just consider the $SU(p)$ version of this model, which for the repulsive
interaction was
first studied by Sutherland\cite{Suth} and insert the
flux via twisted boundary condition. For an attractive interaction
this model has been studied by Takahashi\cite{Taka}.
 The model allows to find
the exact analytical solution in the limit of the strong coupling $V_F/U<<1$,
which is a generalization of the discussed solution for the Hubbard ring
presented above. Generally speaking, in the expansion with
the parameter $\alpha$ the Bethe ansatz equations of the SU(p) model are
decoupled into
a system of $p-1$ dimensionless equations, which describe
$p-1$ fermion reservoirs of "spinons" and a single equation which describes
the charge degrees of freedom.

 There occurs $(p-1)$ fictitious fluxes $\phi_i$ associated with
$p-1$ types of "spinons", which are
relative fluctuations of the charge density between channels.
 In other words,
one may say that there are $M_1, .., M_{p-1}$ spinons, located
on the $p-1$ channels. The "holons", the number of  which  is
 equal to the total number of particles $N$  move in the
 fictitious field created by the $p-1$ sorts of spinons.

The total energy of the multichannel ring in magnetic field
extracted from the Bethe ansatz solution
 has the form:
\begin{eqnarray}
E = \frac{V_F}{L} ( f- \phi_1-\phi_2-... -\phi_{p-1})^2 -
\frac{V_F}{L} \alpha F(\phi_1,\phi_2,... ,\phi_{p-1})
\label{tot-en-m-ch}
\end{eqnarray}
where the
 fictitious fluxes take the values:
$\phi_i = \frac{2\pi K_i}{N}$, and $ K_i = 0,\pm1,\pm2,...$.
The function $F(\phi_1,\phi_2,... ,\phi_{p-1})$ describes the energy
of $p-1$ reservoirs of spinless fermions("spinons"),
 interacting with each other
with a dimesionless coupling.
The form of  $F(\phi_1,\phi_2,... ,\phi_{p-1})$  depends strongly on the
distribution of particles between channels.
 Therefore, there are $p-1$ different
Fermi momenta associated with  the numbers  $M_1, .., M_{p-1}$ where, for
example, $M_i$ is
the number  of particles in the $i-$th channel. For this reason, the
multidimensional
function $F(\phi_1,\phi_2,... ,\phi_{p-1})$  may have cuspoidal
minima when the flux value $\phi_i$  commensurates with the appropriate
Fermi momentum, i.e. with $ k^F_i$.

The presented expression (\ref{tot-en-m-ch}) allows us to get
the following results for the multichannel ring:
analogous to the case of the two channel ring,
there is an enhancement of the persistent current for an intermediate
value of the electron-electron interaction;
2) there are $(p-1)$ fractional periodicities ${M_i}/{N}$,
related to the particle distribution between channels.

The other important point is the dependence of the persistent current
on the number of channels. Note that  $V_F=tN/L$  is related to
the total number of particles $N$ on the ring and therefore
the Fermi velocity in a single channel $v_F$ is equal to $V_F/p$.
That is, the amplitude of the persistent curent is proportional
to $v^2_F p^2/(LU)$ and increases as $p^2$ with the number of channels.
Therefore, with an increase of the number of channels one may
expect a very strong enhancement of the persistent current.
 The disorder, however,  suppresses this  enhancement, but  the amplitude
of the persistent current may  still remain very large and may be compared with
the current of the free electrons. Whether or not this effect can
explain the discrepancy between experiments \cite{22} and \cite{21}
and existing theories
is still unclear.

However, one may suggest that the large amplitude of the current on a single
metallic ring\cite{21} is due to an effect described in the present
paper( due to an large number of channels).
The small value of the current in the Levy et al experiment ($\sim 10^{-2}
v_F/L$) is due to
an average over an ensemble of rings with different number of
particles. Such averaging explains well both the half-flux quantum periodicity
as well as the suppression of the persistent current. In the latter
experiment
the current on the different rings has different orientations
resulting in a compensating effect.
Since the distribution of particles over the channels is, probably, well
controllable, it should be a realizable experiment
to see the predicted
fractional periods.

Recently we became aware of the work of Avishai and Berkovits
\cite{Avis2}, where they also have seen in numerical simulations
of  small disordered rings with 2 and 3 channels an enhancement of the
persistent
current.

I thank Y. Avishai, Alan Luther, Alex Nersesyan,
Minoru Takahashi, Yu Lu, T. Ando, S. Katsumoto, M. Kohmoto,
M. Ueda, K.  Kawarabayashi,  S.M. Manning, F. Assaad
and M. Kohno for useful discussions,
 Ministery of Education, Science and Culture of Japan,
 for support and  ISSP for the hospitality.

\newpage
{\large \bf Figure Captions}\\

\bigskip

{\bf Fig.1}  The behavior of the ground state
 energy  as a function of
flux  $f$ ( the lower parabolic-cusp curve)
and the spinon energy ( indicated by the short horisontal lines)
for 303 electrons with the 101 up spin electrons
at the values $L=1000$ and $U=10$ in the region
 within the half of fundamental flux quantum.
The energy is expressed in
the units $t  10^2$ . The zero energy corresponds to $-520.0 t$.

\bigskip

{\bf Fig.2}  A) The behavior of the holon
 energy   ( the  parabolic-cusp curve) and the spinon energy,
indicated by short horisontal lines
 as a function of
flux  $f$ ($\Phi$) for 303 electrons ($M=101$) at the values $L=20000$ and
$U=10$ in the region within the half of fundamental flux quantum.
 The zero energy corresponds to $-605.771443 t$.
The energy is expressed in
the units $t  10^7$ . B) The behavior of the ground state
 energy  as a function of
flux  $f$.

\bigskip

{\bf Fig.3}  The behavior of the persistent current as a function of
flux  $f$ for 303 electrons at the values $L=20000$ and $U=10$ in the region
of flux within the half fundamental flux quantum. The 101 particles have
up-spin. The maximal current amplitude is equal to $\sim 10 V_F/L$. On the
slope
where  the current  increases, there are perfect 1/303 flux quantum
periodic oscillations.

\end{document}